# Formalising the multidimensional nature of social networks

David Lusseau[1], Louise Barrett[2,3], S. Peter Henzi[2,3]

[1]University of Aberdeen, Institute of Biological and Environmental Sciences, Aberdeen, AB24 2TZ, UK, Email: d.lusseau@abdn.ac.uk, Tel: +44 1224 27 2843; [2]University of Lethbridge, Psychology Department, Lethbridge, Alberta, Canada, T1K 3M4; [3]University of KwaZulu-Natal, School of Psychology, Durban 4041 South Africa

## Abstract

Individuals interact with conspecifics in a number of behavioural contexts or dimensions. Here, we formalise this by considering a social network between $n$ individuals interacting in $b$ behavioural dimensions as a $nxnxb$ multidimensional object. In addition, we propose that the topology of this object is driven by individual needs to reduce uncertainty about the outcomes of interactions in one or more dimension. The proposal grounds social network dynamics and evolution in individual selection processes and allows us to define the uncertainty of the social network as the joint entropy of its constituent interaction networks. In support of these propositions we use simulations and natural 'knock-outs' in a free-ranging baboon troop to show (i) that such an object can display a small-world state and (ii) that, as predicted, changes in interactions after social perturbations lead to a more certain social network, in which the outcomes of interactions are easier for members to predict. This new formalisation of social networks provides a framework within which to predict network dynamics and evolution under the assumption that it is driven by individuals seeking to reduce the uncertainty of their social environment.

**Keywords:** social networks, entropy, evolution

## 1. INTRODUCTION

Social animals benefit in a number of ways by living with others, despite persistent local conflicts of interest [1]. Selection should therefore favour animals that can act in their own interests without at the same time damaging the integrity of their social network [2]. This is an intrinsically difficult task, however, because the ongoing activity and interactions among network members require an ability to predict and respond appropriately to others, given a particular context [3-6]. This greatly increases the number of possible solutions to any particular social situation and means that certainty of outcome necessarily decreases as options multiply. This, in turn, entails acting in ways that can constrain the range of possible outcomes [7-9].

## 2. MULTIDIMENSIONALITY OF SOCIAL NETWORKS

Group members interact with one another in several social or behavioural contexts ('dimensions' hereafter) and the sum of the interaction networks that correspond to these dimensions classically is thought of as the social network to which individuals belong [10-12] (Figure 1). At the same time, however, interactions in one dimension are not independent of interactions in others [13]; two members of a monkey group, for example, can groom each other only to the extent that they can maintain spatial proximity.

This idea of a multidimensional social network, with its associated aggregate reduction of uncertainty, links directly to the key property of a social network, which is its 'small world' nature [14; 15]. In its human context, this small world property is intuitively easy to interpret: even though one may not know all the members of one's social network, it is nevertheless possible to retrieve information from, or transmit it to, any other member. This is because almost all members of a network are linked indirectly through a short chain of intermediary acquaintances. This extensive 'reach' across nodes — indicating the existence of a giant component in the network [16] — thus emerges because individuals can be ascribed a set of identities that combine an array of categories, from the geographic (e.g., nationality, city of residence) to the social (e.g., profession, social group memberships, family ties) and the network can then be searched effectively using these ascriptions [15; 17].

We propose here that the set of identities we can use to categorise individuals is not a property of the individuals themselves (the nodes/vertices in the network), but a property of the relationship between individuals (the edges between individuals). This is because we can only categorise individuals if we interact with them, or gain information about them from their interactions with others. This means that we can think of a social network not as a matrix linking individuals that have a set of identities, but as a set of individuals interacting in a number of behavioural contexts. In the latter situation, the multidimensional 'information' is stored in the interactions (as a set of interaction networks defined by their behavioural context).

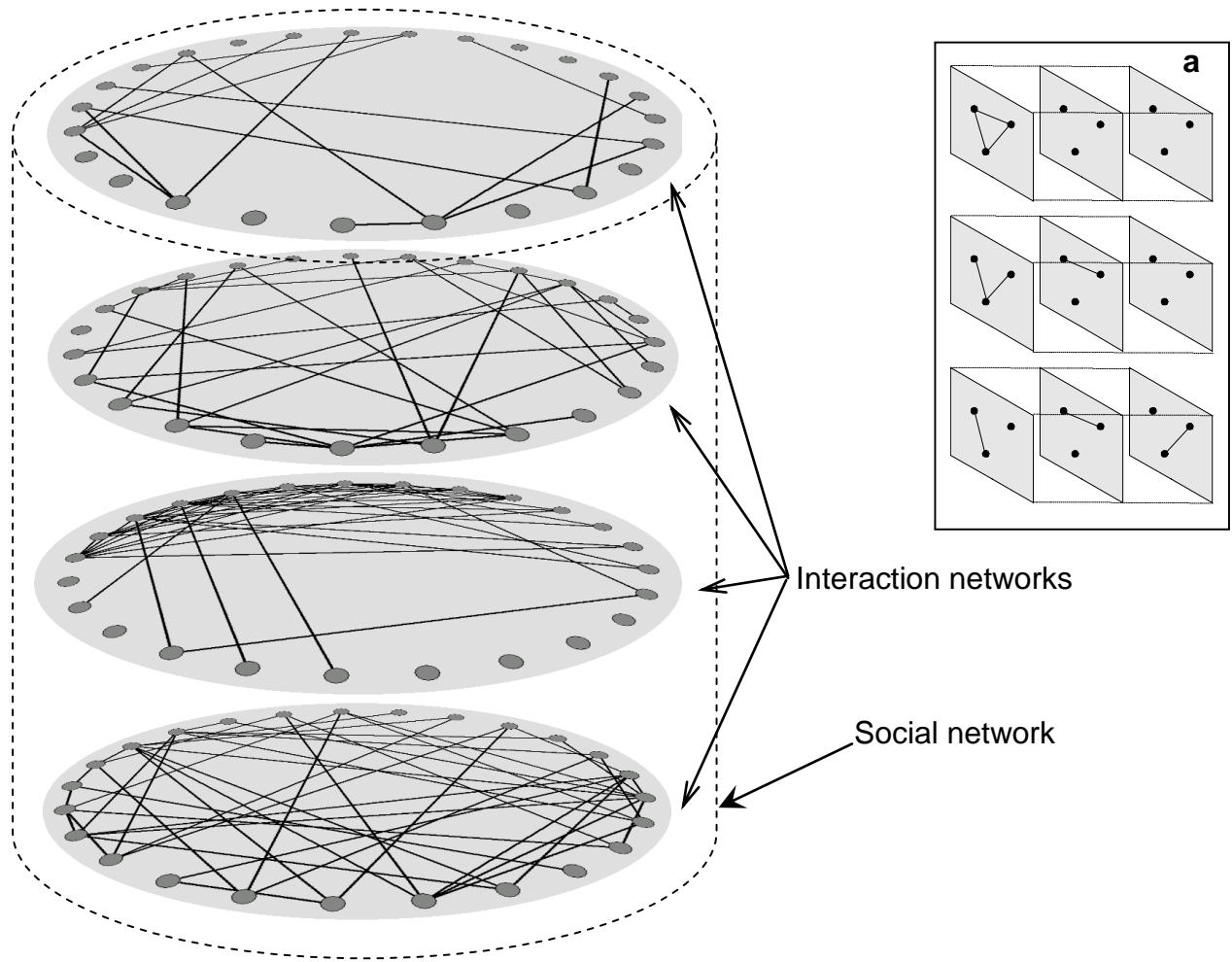

**Figure 1**. A social network is composed of *b* interaction networks between *n* individuals (here n=25). Edges between individuals are defined as interactions in b different behavioural contexts (dimensions). the clustering coefficient of a vertex (individual) is essentially calculated by estimating out of all possible triads in which the vertex is involved how many of those are triangle (all members of the triad connected). When individuals can interact in several behavioural dimensions, triangles and triads can be defined in a number of combinations (**a**: for example, here *b*=3). We can see that in these three cases, if we simply assumed the social network to be the union of the three interaction networks then the clustering coefficient of all three vertices would be 1, when all three examples present different clustering configurations. Our clustering coefficient statistic captures this heterogeneity.

## 3. TOPOLOGICAL CONSTRAINTS REDUCE THE UNCERTAINTY OF SOCIAL ENGAGEMENTS

The idea that uncertainty-reduction drives social interaction is not new and has emerged independently in several disciplines, although it still lacks a formal analytical framework [18-22]. Nevertheless, reducing the inherent complexity of negotiating the social world by moderating the level of uncertainty experienced with respect to others has come to be seen as central to individual fitness [18; 23]. The primary way in which animals can control "social information flow" is to constrain their interactions with others [6]. The increased consistency in the patterns of behaviour so produced can then be captured, in abstract terms, by the topological features characteristic of interaction networks, which have long been known to be associated with patterns of information flow [23; 24]. Given this, we can envisage an interaction network as an information transfer medium, within a particular context, where information, crucially, is defined as a reduction of uncertainty and measured as Shannon entropy [25]. This formalisation does not assume any 'active' information transfer between individuals because it relies on 'structural information' and is therefore broadly applicable [26]. A reliably structured pattern of engagement between pairs of individuals reduces the set of possible interactions in the given interaction network both directly (for the dyad concerned) and indirectly (for their neighbours and their neighbours' neighbours). This is the foundation for the development of, for example, mechanisms explaining the evolution of dominance hierarchies [27] and cooperative behaviour [28]; two of the many potential behavioural contexts in which individuals interact.

Such structural information can be captured using entropic measures because edges in these interaction networks represent the probability that two individuals will engage in a given behavioural context. Shannon extended the thermodynamic concept of entropy to communication following similar principles [25]. The integration of this probabilistic source of information for each individual, over all possible dyadic engagements, then represents a measure of the engagement constraints an individual faces, with the added evolutionary benefits of decreasing the uncertainty of outcomes highlighted earlier. We can then use the Shannon entropy of an interaction network as a measure that integrates this 'constraint as uncertainty-reduction' information over all individuals in the network [6].

Given the multidimensionality of social networks (Section 2), we can then envisage that, in addition to these within-interaction-network constraints, it is also possible to have within-social-network constraints (those between behavioural dimensions) that describe the social network's global efficiency [29]. Although different behavioural dimensions are likely to be differentially constrained [30], their intersection will produce an aggregate reduction of uncertainty that cannot be derived solely from the reduction in each of the component networks because these are not independent of one another. This feature makes it possible to think of a social network, not simply as the sum of its component interaction networks, but as a multidimensional object with emergent properties of its own.

## 4. EVOLUTION OF SOCIAL NETWORKS

When the realized structure of the social network carries benefits, we can expect selection to operate on mechanisms that regulate the tenor of an individual's social conduct [6; 31-33] in ways that will improve the robustness of the multidimensional network structure. Recent experimental work on captive macaques, for example, has illustrated the emergence of particular behavioural mechanisms, such as 'policing', that enhance the resilience of the social network to perturbation, such as that caused by social conflict [13]. Importantly, such policing stabilizes social networks indirectly by facilitating the formation and robustness of "social niches". Niches, in this sense, are the behavioural connections that individuals form in the set of interaction networks in which they participate [13]. Under our premise, these niches are simply the egocentric networks of individuals within a social network; that is, the set of egocentric networks each of which corresponds to an interaction network. This set is not trivial as interaction networks are not independent. The structure of a social network can then be thought of as emerging from the trade-off between the opportunities that an individual encounters and its current needs in each dimension. The structure reduces uncertainty, as in the case of dominance hierarchies [27], because it reduces the range of options and, consequently, the costs of action by group members. Given this, measuring the information content of a social network (as defined in section 3) makes it possible to estimate the degree to which interactions reduce uncertainty. This information represents a measure of the structural constraints that reduce the uncertainty of social engagements for individuals. As this estimate of uncertainty is based on the derived composite structure of the social network, it informs us about the topological state reached by the network as a consequence of trade-offs between the interaction needs of all individuals.

We therefore propose that social networks be considered as multidimensional objects composed of *n* individuals that can interact in *b* behavioural dimensions and that they do so in ways that reduce overall uncertainty. A mathematical representation of a social network in this case is not a matrix of *n* x *n* individuals but a third-order tensor of *n* x *n* individuals x *b* behavioural dimensions. Considering a social network as a higher-order tensor offers new avenues for the formalisation of social structure and dynamics, allowing us to extend Flack et al's earlier consideration of the relationship between networks and social organization [13]. It makes possible the estimation of interaction uncertainties by measuring the social network's Shannon entropy and thereby provides a measure with which to test our central premise that uncertainty reduction drives social network dynamics. From this we can predict:

1. If our proposed theoretical concept is valid we should be able to retrieve characteristic small-world features from our multidimensional networks. That is, social networks, as we formalise them, should retain the high average clustering coefficients that define the cliquish nature of regular networks while also having short chains of intermediary acquaintances (small average path lengths) typical of random networks. Watts and Strogatz [14] used simulations to demonstrate that this small world state emerges between regular and random states as links between individuals in regular networks are randomly rewired. We replicated the Watts and Strogatz simulation procedure in our multidimensional framework and, after deriving new network statistics, tested whether small world features can emerge in multidimensional networks.

2. Perturbations to the social network should increase uncertainty. Evolutionary mechanisms should be in place under our premises so that such perturbations are followed by a corresponding reconfiguration of interaction networks that restructures the topology of the social network minimising entropy. Accordingly, because of the degrees of freedom associated with increasing uncertainty, we expect the least constrained interaction network (the most uncertain) to change the most. Following Flack et al. [13], we test this by taking advantage of natural removals from a chacma baboon (*Papio* *hamadryas ursinus*) social network [34], to assess the effects of the loss of the most dominant female on the structure of spatial and grooming interaction networks.

## 5. SMALL-WORLD PROPERTIES OF MULTIDIMENSIONAL NETWORKS

We replicated the Watts-Strogatz simulations [14] using their random rewiring procedure. Under our formalisation, their study was restricted to the special case of $b$=1, where the tensor is of second-order (i.e. a matrix). We first defined a social network of $n$ individuals by drawing $b$ regular ring binary interaction networks between them with an average degree $k$. That is, in each interaction network, each vertex (individual) was connected to its $k$ nearest neighbours on the regular lattice and an interaction was either present or absent (links were not weighted). Simulations were carried out for a range of network sizes and behavioural dimensions, although we present results only for $b$=3, $n$=1000 vertices (individuals) and $k$=10. As with Watts and Strogatz's simulations [14], our results extend to other social network sizes, provided that $k$>>ln($n$) to ensure that the random network is connected.

Starting with these regular interaction networks, we rewired a proportion p of all edges (interactions) at random ($p$∈[0,1]). This allowed us to modify the social network from one that was regular ($p$=0) to one that was random ($p$=1). When $b$=1, interaction networks can exist in three structural states: regular, random or small world. These are defined using (i) a local measure of cliquishness, the clustering coefficient, which presents the likelihood that individuals to which a network member is connected are also connected to one another and (ii) a global measure of connectivity, the average path length, which is the shortest path one can take to navigate between a given pair of individuals on the network. To define the clustering coefficient and average path length of the third-order tensor, we extended the definition offered by Watts and Strogatz [14], but accounted for the possibility that individuals could be connected in several behavioural dimensions (Figure 1). This extension is trivial for the estimation of average path length in that it only requires the calculation of the average path length on the union of all interaction networks (if $a_{ij}$=1 for at least one interaction network, then $a_{ij}$=1 on the union of all interaction networks). Conversely, the number of possible triadic combinations in a social network increases with $b$ (Figure 1). We therefore extended the clustering coefficient definition to account for the triads and triangles possibly existing between behavioural dimensions, given the elements of the third-order tensor **A**:

$$C_i = \frac{\sum_{j=1}^{n}\sum_{h=1}^{n}\sum_{k=1}^{b}\left[a_{ijk}\cdot\sum_{l=1}^{b}\left[a_{ihl}\cdot\sum_{m=1}^{b}a_{jhm}\right]\right]}{\sum_{j=1}^{n}\sum_{h=1}^{n}\sum_{k=1}^{b}\left[a_{ijk}\cdot\sum_{l=1}^{b}\max(a_{ihl},a_{jhl})\right]} \qquad [1]$$

We then rewired the social network, varying the proportion of edges rewired from 0 to 1. Rewiring was independent for each value of p. We replicated this procedure 20 times and measured the average path length and clustering coefficient of each resulting social network [14].

Our simulations indicate that by varying the proportion of edges randomly rewired in the social network from 0 to 1, a small world state emerges as the network becomes less ordered and before it reaches a random state. For a wide range of $p$, social networks can realize both a short average path length and high clustering coefficient (Figure 2). In addition, we find that $C_{random} \sim \dfrac{bk}{n}$, in agreement with Watts & Strogatz (for $b$=1), and that for a broad range of $p$ we have $C(p) \sim C(0)$ but $L(p) \sim L(1)$, confirming a small world state (Figure 2).

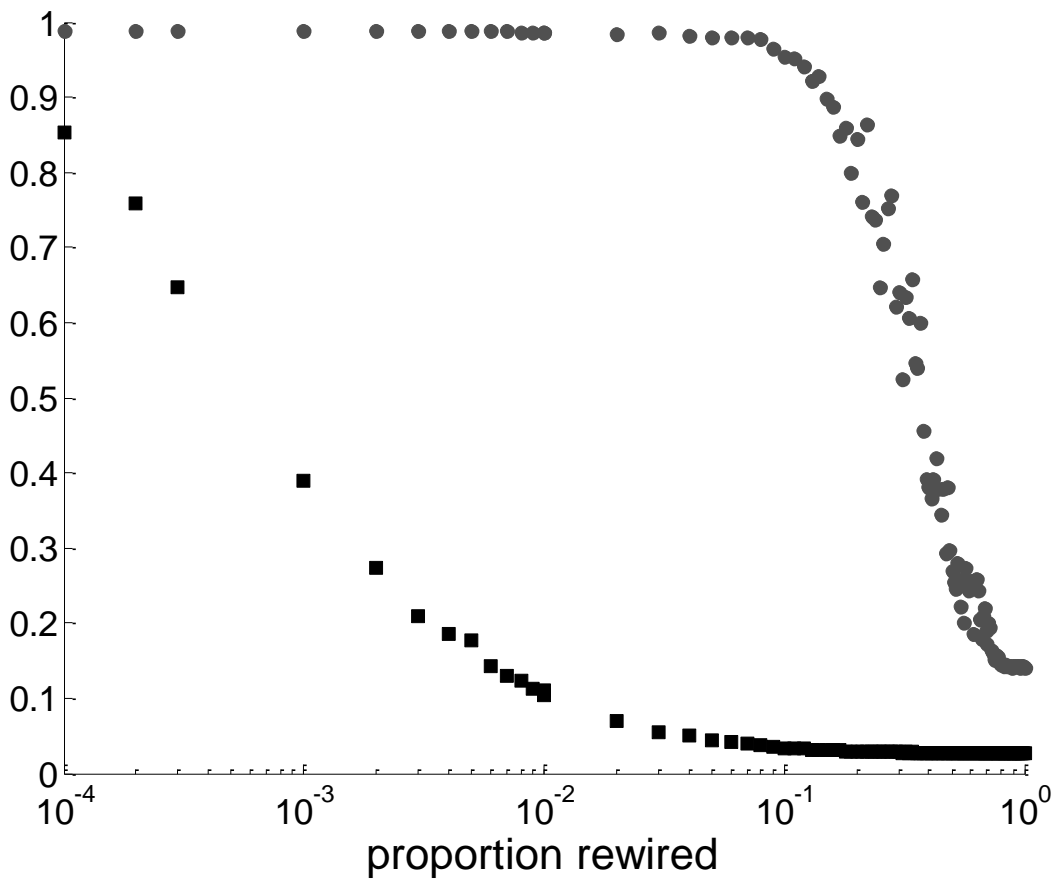

**Figure 2**. Average path length (L) and clustering coefficient (C) for randomly rewired social networks (3-rd order tensor) with the proportion of rewired edges (p) varying from 0 (regular network) to 1 (random network). These values are standardised to their maximum value occurring for p=0. Here n=1000 vertices, k=10 edges per vertex, and b=3 contextual dimensions and the scatterplot represents the average L(p) and C(p) over 20 random simulations.

## 6. CHANGES IN INTERACTION NETWORK TOPOLOGY: AN EMPIRICAL TEST OF THE INTERACTION BETWEEN BEHAVIOURAL DIMENSIONS

We collected aggressive, nearest-neighbour, and grooming observations from March 1997 to October 2006 in our main study troop of baboons in the De Hoop Nature Reserve, South Africa [34]. Data for these analyses come from all adult females (N~12), who were individually recognizable and habituated, and were collected with electronic data loggers while following the troop on foot from dawn to dusk on each sampling day [35]. We used this information to construct a social network for these females, composed of separate agonistic, spatial, and grooming interaction networks.

To test our prediction, we took advantage of unique natural knock-outs [13] to investigate the effects of a disturbance to the agonistic interaction network [13; 30], caused by the disappearance of the dominant female, on the spatial and grooming interactions of other females. Disruptions to the dominance hierarchy will affect agonistic interactions because this structure is present to reduce overall aggression in the troop [27]. We compared the nearest neighbour and grooming networks before and after the death of the dominant female in the troop (Julia). We then compared these changes to control data collected over the same seasons in a different year when troop membership was constant as well as to those observed after the disappearance of a low ranking female (Kersty).

a) **Control (no individuals died)**:
i)      Before: September 2003 to February 2004
ii)     After: March 2004 to August 2004
b) **Dominant female disappearance**:
i)      Before: September 2000 to February 2001
ii)     After: March 2001 to August 2001
c) **Low-ranking female disappearance**:
i)      Before: November 1997 to April 1998
ii)     After: May 1998 to October 1998

This study design [36] allowed us to account for inherent variability in network features due to food availability cycles [34] and perturbation due to changes in troop composition [35]. In addition, we compared the observed variation in network features to a simulated removal condition, in which individuals that disappeared were simply artificially removed from the control network and the network measures were re-calculated.

For each of these six periods, we derived three directed and weighted interaction networks from the agonistic, spatial and grooming data respectively. The network nodes represented individual females and the edges represented the rate at which female $i$ was observed as (a) the nearest neighbour; (b) the aggressor; (c) the groomer of female $j$, given the number of times $i$ was observed. The indices were calculated with a sampling period of one day to address problems of pseudoreplication [37]. The resultant association indices $AI_{ij}$ were not symmetrical and therefore $AI_{ij}$ is not the same as $AI_{ji}$. i.e. the interaction indices are directional. For each interaction network we then calculated the females' clustering coefficients, using the formula for weighted networks [38], which we modified to account for directionality in the data. We could therefore define both an in-clustering coefficient:

$$c_{in_i} = \frac{\sum_j \sum_h AI_{ji} \cdot AI_{hi} \cdot \max(AI_{jh}, AI_{hj})}{\max_{ji}(AI_{ji}) \cdot \sum_j \sum_h AI_{ji} \cdot AI_{hi}}$$
[2]

and an out-clustering coefficient:

$$c_{out_i} = \frac{\sum_j \sum_h AI_{ij} \cdot AI_{ih} \cdot \max(AI_{jh}, AI_{hj})}{\max_{ij}(AI_{ij}) \cdot \sum_j \sum_h AI_{ij} \cdot AI_{ih}}$$
[3]

We bootstrapped observational samples 1000 times with replacement, recalculated behavioural indices for each dyad, and subsequently recalculated clustering coefficients in order to quantify the uncertainty surrounding these network statistics given the data [39]. We used the bootstrapped confidence interval of the difference in mean clustering coefficient between the two levels of treatment to infer the likelihood that clustering coefficients changed significantly [34; 39]. To do so, we carried out paired comparisons, estimating the bootstrapped confidence intervals for each individual and evaluating the significance of this difference. The resulting paired differences differed significantly from zero (p=0.05) if fewer than 50 overlapped [34]. Paired comparisons also blocked the influence of changes in network size after treatment on network statistics [40].

**Table 1.** in- and out- clustering coefficient mean paired differences (in-cc and out-cc, and associated SE) in spatial proximity (Figure 3a,b) for the three treatments (Control, Julia, Kersty) and associated test that the samples come from the same distribution based on two-sample permutation tests for small sample sizes (observed t-statistic based on n samples generated for the permutation distribution ($t_n$) and associated p-values) [41].

|          | **Control**                      | **Julia**                     | **Kersty**     |
|----------|----------------------------------|-------------------------------|----------------|
| **in-cc**| 0.06 (0.014)                     | 0.23 (0.005)                  | -0.07 (0.012)  |
| Control  | -                                | -                             | -              |
| Julia    | $t_{705432}$=11.4, p<0.0001      | -                             | -              |
| Kersty   | $t_{167960}$=7.0, p=0.0003       | $t_{167960}$=25.0, p<0.0001   | -              |
| **out-cc**| 0.07 (0.010)                    | 0.22 (0.006)                  | -0.10 (0.012)  |
| Control  | -                                | -                             | -              |
| Julia    | $t_{705432}$=12.5, p<0.0001      | -                             | -              |
| Kersty   | $t_{167960}$=11.0, p<0.0001      | $t_{167960}$=26.1, p<0.0001   | -              |

After the dominant female disappeared and the agonistic network changed (Julia treatment), all females became significantly more cliquish in their spatial associations (Figure 3a,b). Changes in the agonistic interaction network, however, did not affect the grooming behaviour of female baboons (Figure 3c,d). While significant differences in clustering were also observed in the Kersty treatment, the dominant female knock-out condition was the only level for which *all* individuals changed their behaviour significantly, and the paired differences were significantly greater than the ones observed in the other treatments (Table 1, Figure 3a,b). Prior to the dominant female's death, therefore, female interactions were less conservative, in that a female's nearest neighbours were less likely also to be recorded as one another's nearest neighbours.

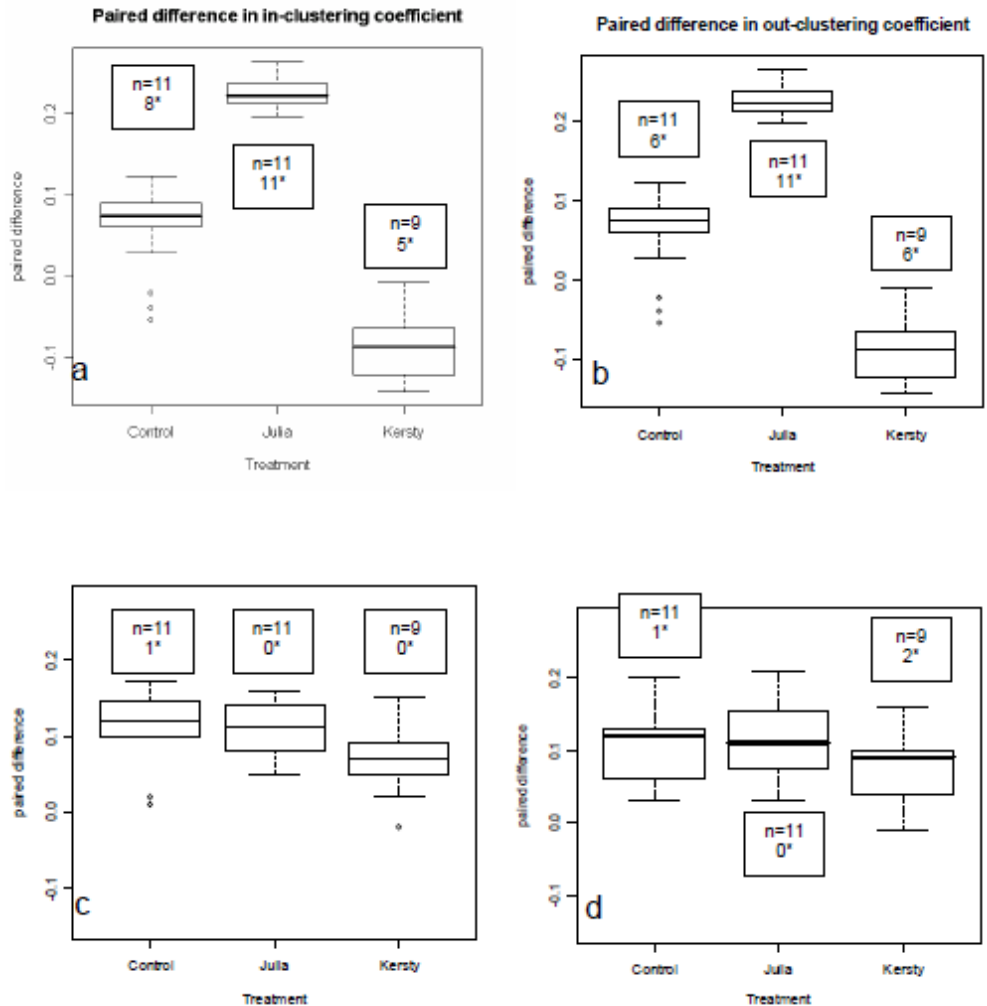

**Figure 3**. Paired differences in in- (a,c) and out- (b,d) clustering coefficients for the nearest neighbour (a,b) and grooming (c,d) directed weighted social networks. Caption above each boxplot provides the number of paired comparisons calculated (n) and the number of individuals that had a significant change in clustering coefficient after the treatment (*).

## 7. UNCERTAINTY AND CONSTRAINTS ON THE SOCIAL NETWORK

Given that we are interested in the capacity of the social network to reduce uncertainty for its members, we estimated the Shannon entropy of this object (Eq. 6) [25]. We first inferred the constraint of each behavioural dimension by estimating the Shannon entropy for each interaction network under each condition. We could then test whether the least constrained dimension (*post-hoc* inference) was the one that changed most in the wake of female disappearance.

We calculated the entropy of the social network as the joint entropy of the interaction networks of which it was composed. We made use of the way the interaction information (I) is formulated to derive the joint entropy (H) of this *b* set of objects. McGill's expansion of the mutual information for more than two variables leads to the interaction information [41]. In our case study, we are dealing with three behavioural dimensions: grooming (*G*), agonism (A), and spatial (*S*) ($\sum_i g_{ij} = \sum_i a_{ij} = \sum_i s_{ij} = 1$), and we therefore have:

$$I(G,A,S)=H(G,A)+H(G,S)+H(A,S)-H(G)-H(A)-H(S)-H(G,A,S) \qquad [4]$$

and, for example,

$$I(G,A,S)=H(G,S)+H(G,A)-H(G)-H(A)-H(S)-H(G|A,S) \qquad [5]$$

Hence the joint entropy of the social network is:

$$H(G,A,S)=H(G,A)+H(G,S)+H(A,S)-H(G)-H(A)-H(S)-I(G,A,S) \qquad [6]$$

where, $H(X)=\sum_{i,j} x_{ij}\log_2 x_{ij}$ , by convention if $x_{ij}=0$ then $h_{ij}=0$.

Our knock-out data, as with Flack et al. [13], indicate that perturbations to the agonistic network lead to more conservative spatial associations. We then expect an associated reduction in the joint entropy of the social network, since increased conservatism should reduce uncertainty. We used jackknife re-sampling, with one individual omitted from the network in turn and the entropy measures recalculated, to derive standard errors for these estimates. We found that perturbations led to social networks in which interactions were less uncertain than they had been prior to the disappearance of females (Figure 4a: 'Julia and 'Kersty treatments). In all cases, the drop in joint entropy was greater than that derived from a simple simulated perturbation (random removal of individuals from the 'before' social networks).

We therefore consider this to be a formalisation of the effects observed in the Flack et al. experiments [13] because a more certain social network corresponds to one in which interactions are more conservative. This is consistent with our hypothesis that social network dynamics processes should be investigated at the social network scale and that we cannot infer expected changes in the social network from an isolated assessment of each interaction network. It also reveals that extrinsic constraints (variation of social structure due to ecological variability between seasons: control treatment) influence the entropy of interaction networks, but not the entropy of the social network as a whole (Figure 4a). The spatial interaction network was the least constrained in all conditions (Figure 4b). This observation, in conjunction with the observed significant changes in interaction networks after the natural removals, agrees with our prediction that the less constrained behavioural dimension should be the one that changes the most.

## 8. DISCUSSION

It is generally accepted that individuals interact with conspecifics in different behavioural dimensions [42] and there is now experimental evidence that these dimensions do not operate in isolation [13]. We developed our formalisation of social networks under the premises that (i) individuals perceive conspecifics as part of their broader ecological landscape [7] and (ii) that behaviours and other traits that reduce uncertainty in that landscape increase their fitness and are therefore adaptive [8]. The trade-offs emerging from contrasting interaction needs for all individuals should then be detected in the 'global information content' of the social network as measured by its joint entropy. We show here that this measure is a meaningful, and biologically relevant, way to describe network dynamics (Figure 4a). It is also a model of social network evolution that can be grounded in individual selection, through the consequences of selection on norms of social reaction in one or more dimensions [32; 33]. This removes the basis for a long-standing reluctance [43] to consider group-level features, such as roles, in the social analysis of anything other than humans. Our results, with those of Flack et al [13], suggest that the direct investigation of structurally imposed constraints on – or regulation of – action can add depth to our understanding of individual social behaviour.

Our simulations demonstrate that small-world properties, a key feature of social networks, can be retrieved from our higher-order formalisation using a new definition and metric for clustering. We defined a new network statistic that helps quantify those changes in social network structure that have previously only been described qualitatively. Variations in the joint entropy of the social network provides a way to track the effects described in Flack et al.'s experiments [13] and offer support for our proposition that social structure dynamics can be understood under the premise of uncertainty reduction. As such, it suggests that applications of our understanding of entropy for open systems can help us guide inferences and predictions for social system dynamics. Under our proposed formalisation of social networks, the concept of social niches is reduced to the definition of egocentric networks in $b$ behavioural dimensions. Such objects are mathematically easier to deal with than earlier proposals [13], and allow us explore how the techniques and metrics that describe egocentric uni-dimensional networks can be extended to characterize the topology of egocentric multidimensional social networks.

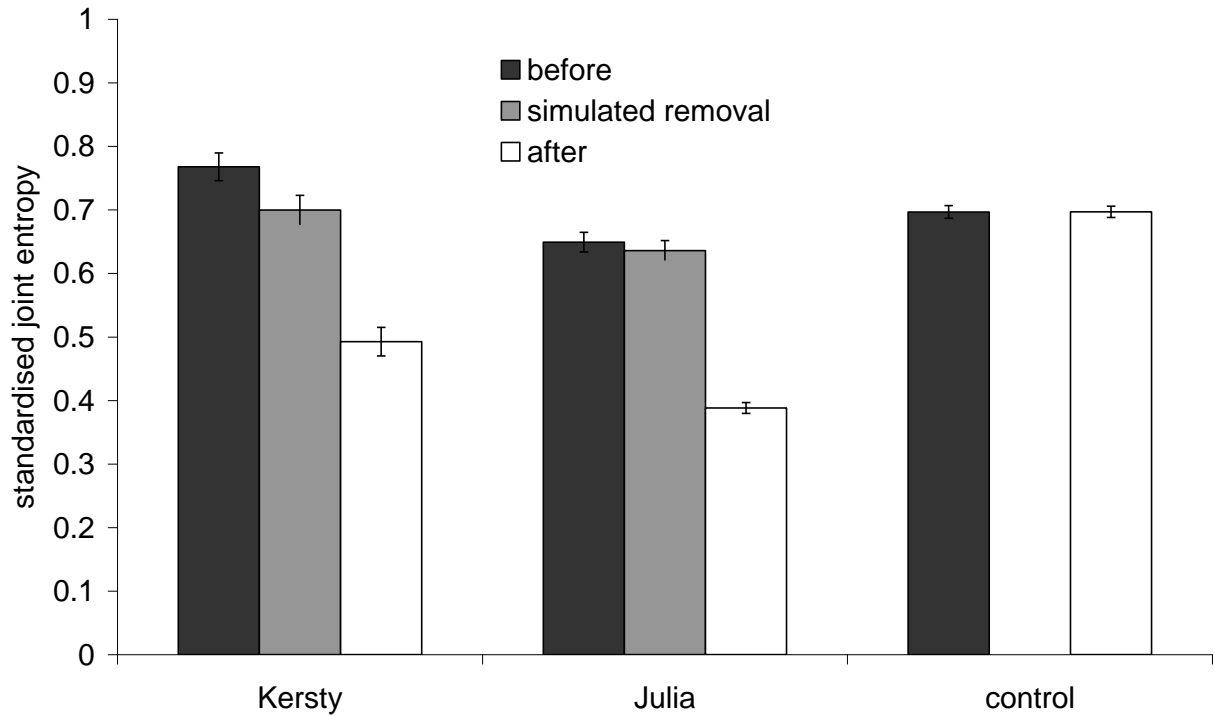

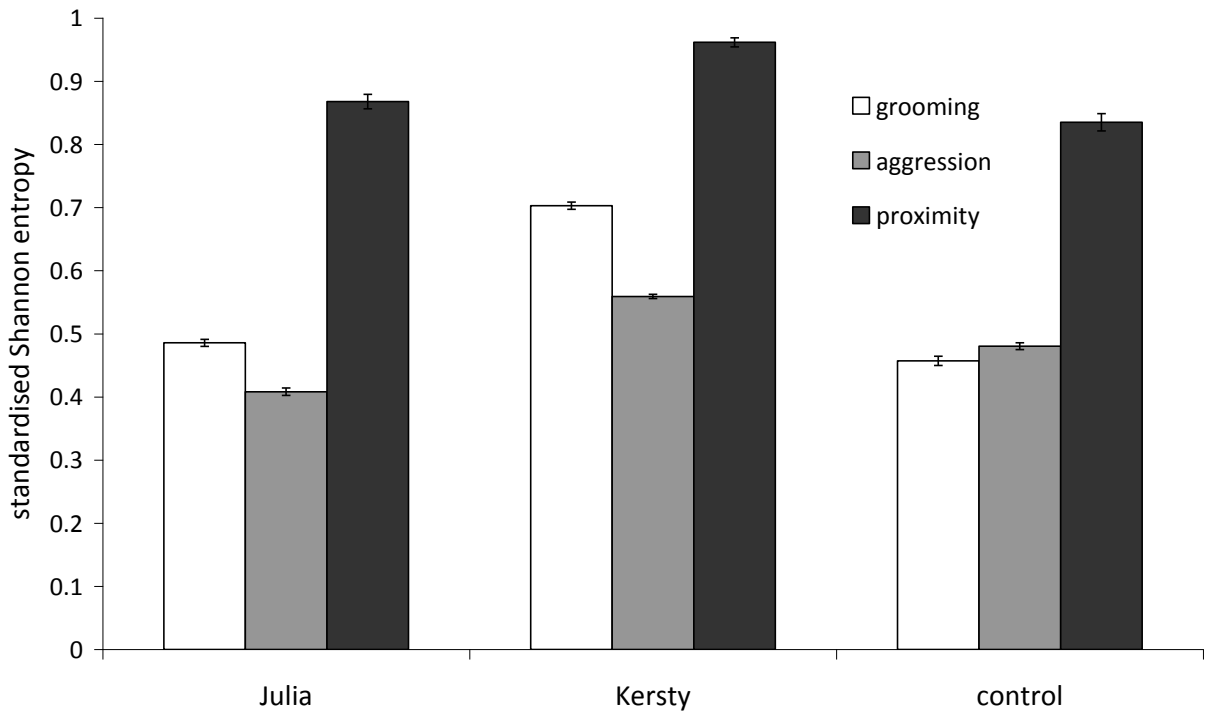

**Figure 4**. Changes in the standardized joint entropy of the social network under the different treatments (a) and the entropy of each interaction network for each treatment 'before' conditions (b). These values range from 0 (highly constrained interactions) to 1 (homogeneous interactions). Error bars are jackknifed standard errors.

Our results, including the qualitative differences observed between the grooming and nearest neighbour networks, we assume driven by the difference in uncertainty between these two behavioural dimensions, also confirm that some individuals can exert a disproportionate influence on social network stability through their actions in particular interaction networks (Figure 4). The use of knock-out designs to assess social niches therefore provides powerful conceptual and methodological tools for probing the forces that create and maintain the social structure of groups. Additional replicates of our treatments would have increased the power of this analysis, however they were not available. Our findings also indicate that, even in the absence of specific mechanisms for conflict-management, dominance stabilizes the social network, promotes network robustness, and increases its information content. Our data allow us to confirm that it is a specific aspect of social structure and not ecological variability or the constraints of captivity that generates these network differences. Finding evidence for this in the wild indicates that it has wide generality and significance for studies of sociality across a broad array of species.

Lastly, if social networks are multidimensional objects, we can hypothesize about the evolution of social network complexity. Under an uncertainty minimization process, the network will reach a $H_{min}$ given $b$, the number of behavioural dimensions, and $n$, the number of individuals (Eq. 6). If we assume that population size (the number of vertices in the network) remains quasi-constant, as for example, when the population is at carrying capacity, then the only way for individuals to reduce their uncertainty further would be to increase $b$. They would, in other words, need to develop new behavioural dimensions in which to interact with others. So the paradigm we present provides a mechanistic process through which the co-evolution of behavioural complexity ($b$) and social complexity ($H_{min}$) can be investigated.

## ACKNOWLEDGEMENTS

DL thanks the Killam trusts for early support for this project. Fieldwork on baboons has been funded by NSRC (Canada) and NRF (South Africa) grants to SPH and LB. We would also like to thank Chris Sutherland, Hal Whitehead, and Janne-Tuomas Seppänen for insightful comments.

## REFERENCES

1. Conradt, L. & Roper, T. J. 2005 Consensus decision making in animals. *Trends in Ecology & Evolution* **20**, 449-456.
2. de Waal, F. B. M. 2000 Primates - A natural heritage of conflict resolution. *Science* **289**, 586-590.
3. Henzi, S. P. & Barrett, L. 2007 Coexistence in female-bonded primate groups. *Advances in the Study of Behavior* **37**, 43-81.
4. Humphrey, N. K. 1976 The social function of intellect. In *Growing points in ethology* (ed. R. A. Hinde), pp. 303-321. Cambridge: Cambridge University Press.

5.      Fowler, J. H. & Christakis, N. A. 2010 Cooperative behavior cascades in human social networks. *Proceedings of the National Academy of Sciences of the United States of America* **107**, 5334-5338.

6.      Eagle, N., Macy, M. & Claxton, R. 2010 Network diversity and economic development. *Science* **328**, 1029-1031.

7.      Barrett, L., Henzi, S. P. & Rendall, D. 2007 Social brains, simple minds: does social complexity really requires cognitive complexity? *Philosophical Transactions of the Royal Society B-Biological Sciences* **362**, 561-575.

8.      Dall, S. R. X., Giraldeau, L. A., Olsson, O., McNamara, J. M. & Stephens, D. W. 2005 Information and its use by animals in evolutionary ecology. *Trends in Ecology & Evolution* **20**, 187-193.

9.      Lima, S. L. & Zollner, P. A. 1996 Towards a behavioral ecology of ecological landscapes. *Trends in Ecology & Evolution* **11**, 131-135.

10.     Adams, B. N. 1967 Interaction theory and the social network. *Sociometry* **30**, 64-78.

11.     Hinde, R. A. 1976 On describing relationships. *Journal of Children Psychology and Psychiatry* **17**, 1-19.

12.     Szell, M., Lambiotte, R. & Thurner, S. 2010 Multirelational organization of large-scale social networks in an online world. *Proceedings of the National Academy of Sciences of the United States of America* **107**, 13636-13641.

13.     Flack, J. C., Girvan, M., de Waal, F. B. M. & Krakauer, D. C. 2006 Policing stabilizes construction of social niches in primates. *Nature* **439**, 426-429.

14.     Watts, D. J. & Strogatz, S. H. 1998 Collective dynamics of 'small-world' networks. *Nature* **393**, 440-442.

15.     Travers, J. & Milgram, S. 1969 An experimental study of the small world problem. *Sociometry* **32**, 425-443.

16.     Newman, M. E. J. & Watts, D. J. 1999 Scaling and percolation in the small-world network model. *Physical Review E* **60**, 7332-7342.

17.     Watts, D. J., Dodds, P. S. & Newman, M. E. J. 2002 Identity and search in social networks. *Science* **296**, 1302-1305.

18.     Danchin, E., Giraldeau, L. A., Valone, T. J. & Wagner, R. H. 2004 Public information: from nosy neighbors to cultural evolution. *Science* **305**, 487-491.

19.     Alchian, A. A. 1950 Uncertainty, evolution, and economic theory. *The Journal of Political Economy* **58**, 211-221.

20.     Tiedens, L. Z. & Linton, S. 2001 Judgement under emotional certainty and uncertainty: the effects of specific emoitions on information processing. *Journal of Personality and Social Psychology* **81**, 973-988.

21.     Gulati, R. 1995 Social structure and alliance formation patterns: a longitudinal analysis. *Administrative Science Quarterly* **40**, 619-652.

22.     Braun, D. P. & Plog, S. 1982 Evolution of "tribal" social networks: theory and prehistoric North American evidence. *American Antiquity* **47**, 504-525.

23.     Granovetter, M. S. 1973 The strength of weak ties. *American Journal of Sociology* **78**, 1360-1380.

24.     Granovetter, M. S. 1983 The strength of weak ties: a network theory revisited. *Sociological Theory* **1**, 201-233.

25. Shannon, C. 1948 *A mathematical theory of communication*. Chicago: University of Illinois Press.
26. Seppänen, J.-T., Forsman, J.-T., Mönkkönen, M. & Thomson, R. L. 2007 Social information use is a process across time, space and ecology, reaching heterospecifics. *Ecology* **88**, 1622-1633.
27. Kaufmann, J. H. 1983 On the definitions and functions of dominance and territoriality. *Biological Reviews of the Cambridge Philosophical Society* **58**, 1-20.
28. Axelrod, R. & Hamilton, W. D. 1981 The evolution of cooperation. *Science* **211**, 1390-1396.
29. Latora, V. & Marchiori, M. 2001 Efficient behavior of small-world networks. *Physical Review Letters* **87**, 198701.
30. Chase, I. D. 1974 Models of hierarchy formation in animal societies. *Behavioral Sciences* **19**, 374-382.
31. Centola, D. 2010 The spread of behavior in an online social network experiment. *Science* **329**, 1194-1197.
32. Kappeler, P. M. & van Schaik, C. P. 2002 Evolution of primate social systems. *International Journal of Primatology* **23**, 707-740.
33. Thierry, B., Aureli, F., Nunn, C. L., Petit, O., Abegg, C. & de Waal, F. B. M. 2008 A comparative study of conflict resolution in macaques: insights into the nature of trait co-variation. *Animal Behaviour* **75**, 847-860.
34. Henzi, S. P., Lusseau, D., Weingrill, T., van Schaik, C. P. & Barrett, L. 2009 Cyclicity in the structure of female baboon social networks. *Behavioral Ecology and Sociobiology* **63**, 1015-1021.
35. Barrett, L., Gaynor, D. & Henzi, S. P. 2002 A dynamic interaction between aggression and grooming among female chacma baboons. *Animal Behaviour* **63**, 1047-1053.
36. Stewart-Oaten, A., Murdoch, W. W. & Parker, K. R. 1986 Environmental impact assessement: Pseudoreplication in time? *Ecology* **67**, 929-940.
37. Whitehead, H. & Dufault, S. 1999 Techniques for analyzing vertebrate social structure using identified individuals: review and recommendations. *Advances in the Study of Behavior* **28**, 33-74.
38. Holme, P., Park, S. M., Kim, B. J. & Edling, C. R. 2004 Korean university life in a network perspective: dynamics of a large affiliation network. *cond-mat/0411634*.
39. Lusseau, D., Whitehead, H. & Gero, S. 2008 Applying network methods to the study of animal social structures. *Animal Behaviour* **75**, 1809-1815.
40. Manly, B. F. J. 1997 *Randomization, bootstrap, and Monte Carlo methods in biology*. New York: Chapman & Hall.
41. McGill, W. J. 1954 Multivariate information transmission. *Psychometrika* **19**, 97-116.
42. Hinde, R. A. 1976 Interactions, relationships and social structure. *Man* **11**, 1-17.
43. Wilson, D. S. 1975 A theory of group selection. *Proceedings of the National Academy of Sciences of the United States of America* **72**, 143-146.
44. Box, G. E. P., Hunter, J. S. & Hunter, W. C. 2005 *Statistics for Experimenters II*. New York: Wiley.